\documentstyle[preprint,aps,floats]{revtex}
\newcommand{\beq}{\begin{equation}}
\newcommand{\eeq}{\end{equation}}
\newcommand{\beqa}{\begin{eqnarray}}
\newcommand{\eeqa}{\end{eqnarray}}

\def\bra{\langle}
\def\ket{\rangle}

\draft
\begin{document}
{\tighten
 
 
\preprint{
\vbox{
      \hbox{December, 2000}
      \hbox{LBNL-47187}}}

\title{On Possibility of Decoherence \\
 in Correlated two Neutral Kaon Systems}
\author{Fumiyo Uchiyama$^{1,2}$}

\address{$^{1}$ Physics Division, Lawrence Berkeley National Laboratory, 
         Berkeley, CA 94720, USA}
\address{$^{2}$ Department of Physics,
     Science University of Tokyo, Noda, Chiba, Japan}


\maketitle

\begin{abstract}
{We point out that decoherence parameters in the correlated two
neutral kaon system breaks the transformation invariance of 
basis and their magnitude are stringently
limited by the experimentally measured magnitudes of CP violation
and of the strangeness non conserving  $\Delta S = \pm 2$ transitions.}
\end{abstract}

\section{introduction}
It has 
been long since the neutral kaons played an important role in the
 history 
of quantum mechanics showing
that the particle mixture\cite{GP} exhibit 
 unusual 
and peculiar 
properties, the oscillating  behaviors in the probabilities 
of finding a $K^0$ (or a $\bar {K^0}$) 
in a beam which is initially pure $\bar K^0$ (or $K^0$), 
as  function of time\cite{PP}. This
property of producing other particles stems 
 from the fact that the strangeness 
eigenstates, 
$K^0$, $\bar {K^0}$ 
are different from eigenstates of Hamiltonian, 
$K_S$ and  $ K_L$.  The fact that   
the inverse of their mass difference is  
the order of one of their life times
made possible for the oscillation to be seen in the laboratory.
The oscillations in the probabilities stem from the interference terms 
 in the particle mixture states coming from quantum mechanical superposition
principle. Based upon these striking quantum mechanical features 
of kaons,
correlated two neutral
 kaons have been used in the proposals for testing 
EPR\cite{EPR}  
 against Quantum Mechanics\cite{fqm}.

Eberhard introduced a decoherence parameter $\zeta$ which parameterizes
 the deviation from Quantum Mechanics 
in his proposals\cite{Eb} on non-locality
experiment.
The experiment involves measurements  of $ K_S $ and/or $ K_L $
 particle states from decays of $\Phi$ mesons; explicitly, 
in the analysis for the proposed experiment,
  the probability of finding two kaons in
states,  $f_1$ and $f_2$,
in the decay products of $\Phi$ mesons, is written as
\beqa\label{phil}
P_{decoh}(f_1 ,f_2) &=& {1 \over 2} [|\bra f_1 |K_{S}\ket 
 \bra f_2 |K_{L}\ket |^{2} +|\bra f_1 |K_{L}\ket \bra f_2 
 |K_{S}\ket |^{2}
 \nonumber \\
&&- (1- \zeta)( \bra f_1 |K_{S}\ket \bra f_2 |K_{L}\ket  
\bra f_1 |K_{L}\ket ^{*}\bra f_2 |K_{S}\ket ^{*} 
\nonumber \\
&&+ \bra f_1 |K_{L}\ket \bra f_2 |K_{S}\ket  \bra f_1 
|K_{S}\ket ^{*}\bra f_2 |K_{L}\ket ^{*})]  
\eeqa
At $\zeta = 0 $,  $P_{decoh}(f_1 ,f_2)$ agrees with the result 
from a quantum mechanical calculation while a
nonzero decoherence parameter indicates an existence of a deviation 
from quantum mechanics. As the extreme case,
 $\zeta =1$ gives zero interference terms
in the above expression. 
Furry\cite{Fr}  in 1936, 
before the discovery of kaons,
  discussed theoretically in detail the degrees of agreement  and 
disagreement between the results of quantum mechanical 
calculations and those to be expected on the assumption 
that a system once freed from dynamical interference
(corresponding to  $\zeta =1$ in the case of  two correlated kaons) 
can be regarded as possessing independently real properties.
Recently the decoherence parameters in
 the interference terms in two neutral kaons,
$K^0$ and $\bar K^0$, are studied \cite{Bert}
in terms of asymmetry in  probabilities for like and unlike strangeness events 
using equation (\ref {phil}). \\
The interference terms in probabilities are 
considered to carry quantum mechanical features  and indeed 
such are  the cases for
the superposition of two waves of one electron passing  different
routes  in space-time and coming back together later,
and  no interference in its superposed wave functions 
means that the system can be treated as  `` classical'', 
namely particle-like. 
Interestingly, it will become clear in this paper
 that the interference terms  in two correlated particle
mixture
do not have this significance.

High energy accelerators
 made it possible to produce
 heavy  
flavored  neutral mesons such as D and B, for which similar
 oscillations could be
observed and
 neutrino
oscillations  among different flavored massive 
neutrinos,  are 
predicted and many  experimental investigation are underway.
Considering these facts,  elucidation 
of the limitation and meaningfulness of decoherence 
(separability in other language) parameters 
in  correlated two particle systems
is in urgent need. It is also important
in the fields of quantum information theory\cite{QIT}.

The purposes of this paper are two:
(1) to show that deduction
 of the magnitude of
interference terms for the case of mixture of particles, even for
the case of
 complete spontaneous factorization($\zeta = 1$) taking place 
 does not imply
that the system is ``separable'', namely ``more classical'', 
contrary to our intuition learned from
the superposition of wave functions in space though it 
 will serve as an indicator
 of deviation from quantum mechanics whenever $\zeta$ is nonzero. 
(2) to show explicitly that the magnitude of 
 decoherence parameters are limited by  the magnitude of
CP violation, and the magnitudes of strangeness 
non-conserving $|\Delta S| = 2 $ transitions, 
and to show  smallness of decoherence parameters
 leads to the necessity for introducing a new 
decoherence parameters  
 for each pair of different eigenstate measurements if an expression
like equation (\ref{phil})  is desirable to be used for analysis.
%
%
%
%
%
\section{formulation}
We use mass and strangeness eigenstates 
$(K_{S}, K_{L})$ and $( K^{0}, \bar K^{0})$  as the two sets of  bases, 
which are more 
frequently used  eigenstates in experiments rather
 than two arbitrarily chosen sets. This particular choice of sets
enable us  to derive the results
by clear and definite arguments
. The latter set is an  
orthogonal bases set while the former set is not.
 It can be easily seen at the end of this section
 that we can develop our arguments  
 more generally ,
using 
any sets of two independent basis vectors. \\
The bases in the two sets are related at  $\underline{all}$  time
\beqa
  |K_{S}\ket  = {N \over {\sqrt 2}} [(1 + \epsilon) |K^{0}\ket  - 
  (1 - \epsilon)|\bar K^{0} \ket ] 
\label{KSLn0-1} \\
  |K_{L}\ket  = {N \over {\sqrt 2}} [(1 + \epsilon)|K^{0}\ket  +
  (1 - \epsilon) |\bar K^{0} \ket ] 
\label{KSLn0-2}
\eeqa
where $\epsilon$ is the CP violation parameter $(|\epsilon|  \sim
10^{-3})$ and
 N is a normalization constant,  ${1 \over \sqrt{(1+|\epsilon|^2)}}$.

We consider the physical situations in which two neutral kaons are 
produced in correlated states of $J^{PC} = 1^{--}$ due to  $\phi$ 
mesons decay. 
At time t after the  $\phi$ decays, the
 two kaon state is expressed in the strangeness basis
$(K^{0}, {\bar K}^{0})$ as 
\beqa
{1 \over \sqrt{ 2}}(|K_{0}(t)\ket _{r}|{\bar K_{0}(t)}\ket _{l} -
 |{\bar K_{0}(t)}\ket _{r} |K_{0}(t)\ket _{l})
\eeqa
where r and l stand for left and right to distinguish the two kaons.
Or similarly in $(K_{S}, K_{L})$, the two kaon state is,
\beqa
{N' \over \sqrt{2}}(|K_{S}(t)\ket _{r} |K_{L}(t)\ket _{l} - 
|K_{L}(t)\ket _{r} 
|K_{S}(t)\ket _{l})
\eeqa
where
$N'= { {( 1 + |\epsilon|^{2})} \over {(1 - \epsilon^{2})}} =
1 + O(|\epsilon|^{2})$
and
$$|K_{S,L}(t)\ket  =  e^{-{i \over 2} {\lambda_{S, L}}} |K_{S,L}(0)\ket $$ 
where 
$\lambda_{S,L}$ are complex mass eigenvalues  of $K_S$ and 
$K_L$ respectively.

The probability of finding an $f_{1}$-type kaon to the right
 and an  $f_{2}$-type kaon to  the left at time t in 
the  $(K_{S}, K_{L})$ basis 
is given by
\beqa\label{bksln0}
P(f_1 ,f_2 ; t) = {1 \over 2}
|\bra f_1 |K_{S}\ket  _{r} \bra f_2 |K_{L}\ket  _{l} -
 \bra f_1 |K_{L}\ket  _{r} \bra f_2 |K_{S}\ket  _{l}|^{2} + O(\epsilon^{2})
\eeqa
where time t is implicit in kaon states and
the correction from the normalization of order ${O(\epsilon^{2}})$ is
separately written.  
 Similarly the
 probability
expressed  in
  $( K^{0}, \bar K^{0})$ bases, is given by 
\beqa \label{bk0}
 P(f_1 ,f_2,t) = {1 \over 2}
 |\bra f_1 |K_{0}\ket \bra f_2 |\bar K_{0} \ket  - \bra f_1 
 |\bar K_{0}\ket 
 \bra f_2 |K_{0}\ket |^{2}
\eeqa
where we have omitted  r and l subscripts with the 
understanding that the first and second states
 in pairwise have r and then  l subscripts respectively.
Here we make further simplification taking $\epsilon =0 $.
This choice makes it clear that the derivation of our conclusions 
is independent of weak interactions.
To make our conclusions applicable to the 
mass eigenstates,$(K_{S}, K_{L})$,
 we will consider corrections later 
by turning on weak interactions.
The corrections are perturbative and order of $\epsilon$.
Substituting $\epsilon = 0$ into equations (\ref{KSLn0-1}, \ref{KSLn0-2}), 
we get two CP eigenstates,
\beqa\label{KSL}
|K_{S} \ket _{\epsilon =0}  = {1\over {\sqrt 2}} [ |K^{0}\ket  - 
|\bar K^{0} \ket ] \nonumber \\
|K_{L} \ket _{\epsilon =0} = {1 \over {\sqrt 2}} [|K^{0}\ket  +
  |\bar K^{0} \ket ] 
\eeqa
which  form another orthogonal basis set.
In this  basis, the two kaons from a $\Phi$ decays  can be expressed
as 
\beqa
{1 \over \sqrt{2}}(|K_{S}(t)\ket _ {\epsilon =0}
|K_{L}(t)\ket _{\epsilon =0} - |K_{L}(t)\ket _{\epsilon =0} 
|K_{S}(t)\ket _{\epsilon =0})
\eeqa
The probability in this basis
is given by
\beqa\label{bksl}
 P(f_1 ,f_2 ; t) =  {1 \over 2}
 |\bra f_1 |K_{S}\ket _{\epsilon =0} \bra f_2 |K_{L}\ket _{\epsilon =0}
 - \bra f_1 |K_{L}\ket _{\epsilon =0} \bra f_2 |K_{S}\ket _{\epsilon =0}|^{2} 
\eeqa
The two sets of  orthogonal bases to be used from  now on are
, $(K_{S}, K_{L}) _{\epsilon =0}$ and $( K^{0}, \bar K^{0})$ which
are  eigenstates of CP and Strangeness respectively.
They are eigenstates
of the Hamiltonian,
$$ H = H_{st} + H_{em} $$
where $ H_{st}$  and $ H_{em}$ are strong and 
electro-magnetic interaction hamiltonians respectively. 
Hamiltonian, strangeness and CP  operators satisfy commutation relations;
$$ [H, CP] = 0 ; \qquad  [H, S] = 0 ; \qquad  [S, CP] \ne 0 $$

As it is rather annoying to keep the subscripts $ >_{\epsilon =0}$, 
unless otherwise stated, 
we use  the same notations, $K_S$ and  $K_L$, for CP+ and CP-
eigenstates  without subscript  $_{\epsilon =0}$ now on.

Expanding the terms in the equation (\ref{bksl}), the probability 
is given as
\beqa\label{expan}
 P(f_1 ,f_2,t) &=& {1 \over 2} (
 |\bra f_1 |K_{S}\ket \bra f_2 |K_{L}\ket |^{2} +|\bra f_1 |K_{L}
 \ket \bra f_2 |K_{S}\ket |^{2} 
 \nonumber  \\
 && -  \bra f_1 |K_{S}\ket \bra f_2 |K_{L}\ket 
 \bra f_1 |K_{L}\ket ^{*}\bra f_2 |K_{S}\ket ^{*}  
 \nonumber \\
 && -\bra f_1 |K_{L}\ket \bra f_2 |K_{S}\ket  
 \bra f_1 |K_{S}\ket ^{*}\bra f_2 |K_{L}\ket ^{*})  
\eeqa
Or equivalently,
\beqa
 P(f_1 ,f_2; t) &=&  {1 \over 2} [|\bra f_1 |K_{S}\ket \bra f_2
 |K_{L}\ket |^{2} +|\bra f_1 |K_{L}\ket \bra f_2 |K_{S}\ket |^{2} 
 \nonumber \\
 && - 2 Re(\bra f_1 |K_{S}\ket \bra f_2 |K_{L}\ket  \bra f_1 |K_{L}
 \ket ^{*}\bra f_2 |K_{S}\ket ^{*})]
\eeqa
To transform the probability to the one  expressed in  the set of bases
 ($ K_{0}, \bar K_{0})$ , we substitute (\ref{KSL}) 
into
(\ref{expan}).  For the first term in (\ref{expan}), we obtain 
by direct substitution,
\beqa\label{first}
 |\bra f_1 |K_{S}\ket \bra f_2 |K_{L}\ket |^{2}
 &=& {1 \over 2}[ \bra f_1 |K_{0}\ket \bra f_2 |K_{0}\ket  -
  \bra f_1 |{\bar K_{0}}\ket \bra f_2 |{\bar K_{0}}\ket  
 +  \bra f_1 |K_{0}\ket \bra f_2 |{\bar K_{0}}\ket  - 
 \bra f_1 |{\bar K_{0}}\ket \bra f_2 |K_{0}\ket ]  
 \nonumber  \\
 && \cdot [ 
 \bra f_1 |K_{0}\ket ^{*} \bra f_2 |K_{0}\ket ^{*} - 
 \bra f_1 |{\bar K_{0}}\ket ^{*}\bra f_2 |{\bar K_{0}}\ket ^{*} + 
 \bra f_1 |K_{0}\ket ^{*}\bra f_2 |{\bar K_{0}}\ket ^{*} \nonumber \\
 &&-  \bra f_1 |{\bar K_{0}}\ket ^{*}\bra f_2 |K_{0}\ket ^{*}] 
\eeqa
In order to make calculation transparent, we define the 
following quantities; 
\beqa
 A \equiv \bra f_1|K_{0}\ket \bra  f_{2}|K_{0}\ket  -  \bra f_1
 |{\bar K}_{0}\ket \bra  f_{2}|{\bar K}_{0}\ket  \\
 B \equiv \bra f_1|K_{0}\ket \bra  f_{2}|{\bar K}_{0}\ket  - 
 \bra  f_{1}|{\bar K}_{0}\ket \bra f_{2}|K_{0}\ket 
\eeqa
Note that A has amplitudes of $\Delta S = 2$ and  $\Delta S = -2$ 
transitions while B is strangeness conserving amplitudes. 
Then equation (\ref{first}) becomes
\beqa\label{base1}
 |\bra f_1 |K_{S}\ket \bra f_2 |K_{L}\ket |^{2}  = {1 \over 4} 
 ( A + B )( A^{*} + B^{*})
\eeqa
Similar substitutions into  the rest of terms in equation (\ref{expan})
in this  bases yields
\beqa\label{base2}
  |\bra f_1 |K_{L}\ket \bra f_2 |K_{S}\ket |^{2} = {1 \over 4} 
  ( A - B )( A^{*} - B^{*})  \nonumber\\
  \bra f_1 |K_{S}\ket \bra f_2 |K_{L}\ket  \bra f_1
  |K_{L}\ket ^{*}\bra f_2 |K_{S}\ket ^{*} = {1 \over 4} 
  ( A + B )( A^{*} - B^{*}) 
  \nonumber\\
  \bra f_1 |K_{L}\ket \bra f_2 |K_{S}\ket  \bra f_1 
  |K_{S}\ket ^{*}\bra f_2 |K_{L}\ket ^{*} = {1 \over 4} 
  ( A - B )( A^{*} + B^{*}) 
\eeqa
Summing all these terms  up, we obtain for transformed probability
in  ($ K_{0}, \bar K_{0})$ bases:
\beqa
P(f_1 ,f_2, t) = {1 \over 2} B B^{*} ={1 \over 2}|B|^2
\eeqa
which is nothing but the equation(\ref{bk0}), the expression 
 in the 
bases of $( K^{0},{ \bar K}^{0})$ showing the invariance of the probability
 $P(f_1,f_2; t)$
under base-transformations.
%
%
%
%
%
\section{Introduction of decoherence parameter}
Following Eberhard\cite{Eb}, let's assume that
 the physical processes of a particular interest
 might deviate from QM for some unknown 
reasons and that one of the conceivable parameterization for the probability 
in a particular experiment involving
 measurement on  pairs  of kaons in mass eigenstates  
 $K_{S}$ and $ K_{L}$ is given by a modification of
 the interference term as
\beqa\label{decoh}
P_{decoh}(f_1 ,f_2;t) &=& {1 \over 2} [|\bra f_1 |K_{S}\ket \bra f_2
 |K_{L}\ket |^{2} +|\bra f_1 |K_{L}\ket \bra f_2 |K_{S}\ket |^{2} 
\nonumber \\
&& - (1- \zeta)( \bra f_1 |K_{S}\ket \bra f_2 |K_{L}\ket  
\bra f_1 |K_{L}\ket ^{*}\bra f_2 |K_{S}\ket ^{*}  
\nonumber \\
&& + \bra f_1 |K_{L}\ket \bra f_2 |K_{S}\ket  \bra f_1
 |K_{S}\ket ^{*}\bra f_2 |K_{L}\ket ^{*}) ] 
\eeqa
where $\zeta$ is called decoherence parameter\cite{Eb}.
We rewrite $P_{decoh}(f_1 ,f_2; t)$ above in the  
$( K^{0},{ \bar K}^{0})$ basis.

  Following the steps we took in the previous section for
transforming bases and  using equations (\ref{base1})
 and (\ref{base2}), 
we obtain
 \beqa\label{dek}
 P_{decoh}(f_1 , f_2; t) = {1 \over 2} B B^{*} + ({  \zeta \over 2})
 ( A A^{*} -  B B^{*} )
 \eeqa
which can be explicitly written down in  the  
$( K^{0},{ \bar K}^{0})$ basis as
 \beqa\label{decoh0}
  P_{decoh}(f_1 ,f_2;t) &=& {1 \over 2} [|\bra f_1 |K_{0}\ket 
  \bra f_2 |{\bar K}_{0}\ket |^{2} +  
  |\bra f_1 |{\bar K}_{0}\ket \bra f_2 |K_{0}\ket |^{2} 
 \nonumber \\
  && - ( 1 - \zeta ) (\bra f_1 |K_{0}\ket \bra f_2 |{\bar K}_{0}\ket  
  \bra f_1 |{\bar K}_{0}\ket ^{*}
  {\bra f_2 |K_{0}\ket }^{*} 
 \nonumber \\
  && + \bra f_1 |{\bar K}_{0}\ket \bra f_2 |K_{0}\ket  
  {\bra f_1 |{K}_{0}\ket }^{*} {\bra f_2 |\bar K_{0}\ket }^{*}) 
  +  \zeta \cdot T_{extra} \;]   
 \eeqa
$\zeta = 0$ reproduces the eq.(\ref{bk0}) as we expect.
Because of the nonzero term, $ T_{extra}$, 
the form of ${ P_{decoh}}(f_1 ,f_2; t)
$ becomes no-invariant under the
transformation of bases.
There is no apriori reason for invariance of the form of 
$P_{decoh}(f_1 ,f_2;t)$  under base transformations .
(However we will see that the invariance requirement is important
for spin -correlated two lepton case at the end of section IV.)
The explicit terms in   $ T_{extra}$ are
\beqa\label{dp}
  && { 1 \over 2}(|\bra f_1 |K_{0}\ket \bra f_2 |{\ K_{0}}\ket |^{2} 
  +|\bra f_1 |{\bar K}_{0}\ket \bra f_2 |{\bar K}_{0}\ket |^{2}  
\nonumber \\
  && -\bra f_1 |K_{0}\ket \bra f_2 | K_{0}\ket  \bra f_1|
 {\bar K}_{0}\ket ^{*}
  \bra f_2| {\bar K}_{0}\ket ^{*} 
  - \bra f_1 |K_{0}\ket ^{*}\bra f_2 |{ K}_{0}\ket^{*}\bra f_1| 
  {\bar K}_{0}\ket \bra f_2| {\bar K}_{0}\ket  
\nonumber \\  
  && -|\bra f_1 |K_{0}\ket \bra f_2 |{ \bar K_{0}}\ket |^{2} 
  -|\bra f_1 |{\bar K}_{0}\ket \bra f_2 |{ K}_{0}\ket |^{2}  
\nonumber \\
  && -  \bra f_1 |K_{0}\ket \bra f_2 |{\bar K}_{0}\ket  
  \bra f_1| {\bar K}_{0}\ket ^{*}
  \bra f_2| { K}_{0}\ket ^{*}  
  -  \bra f_1 |\bar K_{0}\ket \bra f_2 |{ K}_{0}\ket  
  \bra f_1| {K}_{0}\ket ^{*}
  \bra f_2| {\bar K}_{0}\ket ^{*})
\eeqa
where the terms in the  first two lines of the equation come from $|A|^2$ 
and the rest comes from 
$|B|^2$ and the counter terms to cancel the terms added to make the 
desired expression of $\zeta$. 
When $\zeta = 1$,  the obvious
interference terms [ the terms proportional to $(1- \zeta)$]
 in the two probabilities expressed 
in the bases $(K_{S}, K_{L})$, eq.(\ref{decoh}),
and  $(K^{0}, {\bar K}^{0})$, eq.(\ref{decoh0}) respectively  vanish.
But the interference terms in the $ T_{extra}$
in the bases $(K_{S}, K_{L})$, eq.(\ref{decoh0}), 
the last line of $ T_{extra}$, equation(\ref{dp}), are nonzero and become
the maximum.
The first two terms in  equation (\ref{dp}) of $T_{extra}$
 \beqa\label{vk0}
 {\zeta \over 4} (|\bra f_1 |K_{0}\ket \bra f_2 |{\ K_{0}}\ket |^{2} 
  +|\bra f_1 |{\bar K}_{0}\ket \bra f_2 |{\bar K}_{0}\ket |^{2}) 
 \eeqa
 cause $\Delta S= 2$
and  $\Delta S= -2$ transitions respectively in the probability.
We emphasize here that $|\Delta S| = 2$
transitions of order $\zeta$ could occur for $\zeta \ne 0$
even if  the weak interactions has been turned off. 
Nonzero $\zeta$ breaks invariance of the probability under 
 transformation of basis; namely the freedom of choice of the quantum axis
 is violated by a
 nonzero $\zeta$. This may not sound strange because $\zeta \ne
0$ means after all that there exists a deviation from quantum mechanics
upon  which transformation invariance of basis is based upon. 
Therefore we could not  disqualify on this ground  equation
 (\ref{decoh}) 
as a suggestive ways to deviate from  the results obtained 
by quantum mechanics.

For a different experiment involving measurements on pairs of
 kaons
in eigenstates  of strangeness, 
 $(K_{0}, \bar K_{0})$,
one may think to introduce a new decoherence parameter ${\zeta'}$ 
for  $(K_{0}, \bar K_{0})$ bases 
 exactly in 
the same form for $(K_{S}, K_{L})$ bases as
 \beqa\label{decoh1}
 P_{decoh}(f_1 ,f_2; t) 
 &=& {1 \over 2}[|\bra f_1 |K_{0}\ket \bra f_2 |{\bar K}_{0}\ket |^{2} +
 |\bra f_1 |{\bar K}_{0}\ket \bra f_2 |K_{0}\ket |^{2} 
 \nonumber \\
 && - 2 ( 1 - {{\zeta'}} ) Re(\bra f_1 |K_{0}\ket \bra f_2 
 |{\bar K}_{0}\ket  \bra f_1 |{\bar K}_{0}\ket ^{*}\bra f_2 |K_{0}\ket ^{*})]
 \eeqa
where $\zeta'$ = 0 agrees to the result from quantum mechanics and
 $\zeta'$ = 1 implies no interference term as before. \\
The relationship between the two decoherence parameters, 
$\zeta$ and ${\zeta'}$ can be obtained
 by equating 
(\ref{decoh0}) 
and (\ref{decoh1}). That is;
\beqa\label{zetaap}
{{\zeta'}} = {\zeta} \cdot [ 1 + {1 \over 2} {T_{extra} \over 
 Re(\bra f_1 |K_{0}\ket \bra f_2 |{\bar K}_{0}\ket  \bra f_1 |{\bar
 K}_{0}\ket ^{*}\bra f_2 |K_{0}\ket ^{*})}]
\eeqa
To see the magnitude of $\zeta'$,  let us take
 $f_1 =\gamma( |K_0\ket  + \eta |\bar K_0\ket ) $ and  $f_2 =\gamma(
|{\bar K}_0\ket  - \eta |K_0\ket ) $  where $ \eta$ is a small number  as
 an extreme example,
and  $\gamma$ is a normalization. 
  The denominator in equation(\ref{zetaap}) 
becomes as small as  $\sim |\eta|^2$
while the numerator has a finite value, 
$\sim {1 \over 2}$ meaning a large ${\zeta'}$.
This indicates that once $\zeta$  is nonzero, 
small deviation from quantum mechanics
does not necessarily imply a small decoherence parameter in other basis
; more explicitly, an improper
usage of the  expression of equation(\ref{decoh}), which is
 written for the analysis for the experiments involving measurements of
eigenstates, $K_S, K_L$ will cause a large  decoherence parameter
in agreement with the results of reference \cite{Bert} 
in which the base-dependence of $\zeta$
is pointed out.
Regardless, both decoherence
 parameters are severely constrained  as shown in 
the next section
and the expressions
such as  equation (\ref{decoh}) and 
 equation (\ref{decoh1})  can be used
 only for analysis on  high precision measurements.


\section{Upper bounds for the Magnitudes of decoherence parameters}
 $\zeta$ induces
 nonzero probability for strangeness non-conserving,
 $\Delta S= \pm 2$ transitions  of  order  $\zeta$
 as shown in the first two terms in  $ T_{extra}$ of equation(\ref{dp})
 which is;
\beqa\
 {\zeta \over 4} (|\bra f_1 |K_{0}\ket \bra f_2 |{\ K_{0}}\ket |^{2} 
  +|\bra f_1 |{\bar K}_{0}\ket \bra f_2 |{\bar K}_{0}\ket |^{2}) 
\eeqa
This  remains true at any arbitrary time t,   
as long as  two sets of 
 bases are related to each other by equations  
(\ref{KSL}).
To our best knowledge, the  $|\Delta S| =  2$ transitions are 
 limited by order of the weak interaction.
The $|\Delta S|$ = 2 transition probability is proportional
to $\Delta m (\sim 10^{-6}ev)$ 
which is the mass difference of $K_L$ and $K_S$ and can 
be estimated to be 
 $\sim  {\Delta m \over  m_{c}}$ using charm meson mass,  $m_{c}$.   
 Therefore the magnitudes of
 $\zeta$ is limited from the experiment as

$${{\zeta} \over 4 } \quad \le  \quad \sim  {\Delta m \over  m_{c}}
\quad \sim 10^{-12}$$  
The corrections of weak interactions on the results may
 be estimated as follows:
The eigenstates of effective hamiltonian 
$$H_{ef} = H_{st} + H_{em} + H_{weak} $$ 
are give  in equations (\ref{KSLn0-1}, \ref{KSLn0-2}). 
Expanding the states in terms of $\epsilon$,
we obtain
\beqa 
 |K_{S}\ket  = {N \over {\sqrt 2}} [(1 + \epsilon) |K^{0}\ket  - 
 (1 - \epsilon)|\bar K^{0} \ket ] = {1 \over
 {\sqrt 2}}[|K_{S}\ket _{\epsilon=0} +
 \epsilon |K_{L}\ket _{\epsilon=0}  + O(\epsilon^{2})] \nonumber \\
 |K_{L}\ket  = {N \over {\sqrt 2}} [(1 + \epsilon)|K^{0}\ket  +
 (1 - \epsilon) |\bar K^{0} \ket ] ={1 \over {\sqrt 2}}[|K_{L}
 \ket _{\epsilon=0} +
 \epsilon |K_{S}\ket _{\epsilon=0} +  O(\epsilon^{2})]
\eeqa
Therefore the $1^{--}$ state of two neutral kaons
can be written in terms of CP eigenstates as
\beqa\label{masscp}
 (|K_{S}(t)\ket   |K_{L}(t)\ket  - |K_{L}(t)\ket 
|K_{S}(t)\ket ) &=& (|K_{S}(t)\ket  _{\epsilon=0} |K_{L}(t)\ket 
 _{\epsilon=0} 
- |K_{L}(t)\ket  _{\epsilon=0} 
|K_{S}(t)\ket  _{\epsilon=0} ) \nonumber \\
&& +  O(\epsilon^{2})terms
\eeqa
where $O(\epsilon^{2})terms$ represents the sum of  products
of combinations of  $|K_{S}(t)\ket  _{\epsilon=0}
 $  and  $|K_{L}(t)\ket  _{\epsilon=0}$ of order of  $O(\epsilon^{2})$ 
and higher.
Among them, there will be terms such as  
$|K_{S}(t)\ket  _{\epsilon=0}|K_{S}(t)\ket  _{\epsilon=0}$,
and $|K_{L}(t)\ket  _{\epsilon=0}|K_{L}(t)\ket  _{\epsilon=0}$ 
 which are CP violating, CP=+, states .
Note there is no correction of order  $\epsilon$ and
 the corrections
to the equations from  (\ref{decoh}) to  (\ref{dp}) are order of $\epsilon^2$
which can be neglected when we are taking effects of order  $\epsilon$ 
into accounts.
Therefore the upper limits for the magnitudes of
 decoherence parameters for the basis of effective hamiltonian
$(K_{s}, K_{L})$ of 
equation (\ref{phil})is
$$ \zeta \quad \le \sim O(|\epsilon|^2) $$
 Therefore we can safely state
$${{\zeta} \over 4 } \quad \le \quad  |\epsilon| $$  
Similarly, if equation (\ref{decoh1}) is used for
analysis on data from  measurements of strangeness eigenstates,
we inevitably end up  introducing CP violating
transitions terms of order  $\zeta'$;
$$ {{\zeta'} \over 4} (|\bra f_1 |K_{S}\ket \bra f_2 |{\ K_{S}}\ket |^{2} 
+|\bra f_1 |{K}_{L}\ket \bra f_2 |{K}_{L}\ket |^{2})$$
The derivation of this two terms is trivial 
because  the mathematical procedures are exactly  
reversed but identical
for the transformation
from the strangeness basis $(K^{0}, \bar K^{0})$ to CP  basis
$(K_{S}, K_{L})_\epsilon=0$.
As CP violating transitions are bounded by order of $\epsilon$, 
we get
\beqa
{{\zeta'} \over 4} \quad \le \quad |\epsilon| 
\eeqa
Therefore conservatively the  coherence parameters are limited by
\beqa
{{\zeta'}}, \quad {\zeta} \quad \le \quad  4|\epsilon| 
\eeqa
The stringent limit on the decoherence parameters limits
the usage of equations such as  (\ref{decoh})
and (\ref{decoh1}) in the data analysis. Inversely,
a large decoherence parameter
 is a signal for 
an erroneous use of formulation or mistakes made in analysis.

The situation discussed above
can be understood more clearly when we take an example in the
more familiar case of spin  
$ 1 \over 2$ particles, say  two correlated electrons with total spin 0.
The choice of quantization  axis, can be chosen from any one of
axis, x, y, or z. It 
 is free for user's
choice and the contents of physics must be independent of 
the choice one makes. 
Total 
angular momentum is conserved and there has been
 no violation of this conservation
observed so far. We imagine a situation in which
two electrons are moving apart on y axis in opposite direction. 
Two sets of bases may be in z and x axis and their spin
states are related by
\beqa\label{spinSL}
|x\uparrow\ket  = {1 \over {\sqrt 2}} (|z\uparrow\ket  - 
|z \downarrow\ket ) 
\nonumber \\
|x\downarrow\ket  = {1 \over {\sqrt 2}} 
(|z\uparrow\ket  + |z \downarrow\ket ) 
\eeqa
which are analogous to equation (\ref{KSL}):
If we make substitutions 
\beqa\label{coresp}
|x\uparrow\ket   \rightleftharpoons  |K_{S}\ket  \nonumber \\
|x\downarrow\ket \rightleftharpoons  |K_{L}\ket  \nonumber \\ 
|z\uparrow\ket   \rightleftharpoons  |K^{0}\ket   \nonumber \\ 
|z\downarrow\ket \rightleftharpoons  |\bar K^{0}\ket 
\eeqa 
the analogy is clear.
There is no time-dependence in the probability unless we choose 
a time dependent decoherence parameter. 
The probability for finding one of electron spin up and the other down is
\beqa\label{spin} 
P_{spin}(f_1 ,f_2;t) &=& {1 \over 2}[|\bra f_1 |x\uparrow
\ket \bra f_2 |
x\downarrow\ket |^{2} 
+|\bra f_1 |x\downarrow\ket \bra f_2 
|x\uparrow\ket |^{2} 
\nonumber \\
&& - (1- \xi)(\bra f_1 |x\uparrow\ket \bra f_2 
|x\uparrow\ket \bra f_1 |x\downarrow\ket ^{*}\bra f_2 
|x\downarrow\ket  ^{*}
 \nonumber \\
&& + \bra f_1 |x\uparrow\ket ^{*}\bra f_2 
|x\uparrow\ket ^{*}\bra f_1 |x\downarrow\ket \bra f_2 
|x\downarrow\ket )]  
\eeqa
where we introduced  a decoherence parameter
 $\xi $  and  $\xi = 0$ gives the prediction of probability
for two electron's spins, $f_1$
and $f_2$ in  quantum mechanics. 
By four substitutions in equation (\ref{coresp}),
we see that
at $\xi = 0$ the probability is invariant under base transformations
; namely  
changing x quantization axis to z quantization axis, we can show
by substitutions
that the form of 
$P_{spin}(f_1 ,f_2; t)$ in z-basis  does not change.
For none zero $\xi$, by  going through the same calculation 
performed for kaons or again by substitutions of (\ref{coresp}),
we obtain
\beqa\label{spinz} 
P_{spin}(f_1 ,f_2;t) &=& {1 \over 2}[|\bra f_1 |z\uparrow
\ket \bra f_2 |
z\downarrow\ket |^{2} 
+|\bra f_1 |z\downarrow\ket \bra f_2 
|z\uparrow\ket |^{2} 
\nonumber \\
&& - (1- \xi)(\bra f_1 |z\uparrow\ket \bra f_2 
|z\uparrow\ket \bra f_1 |z\downarrow\ket ^{*}\bra f_2 
|z\downarrow\ket  ^{*}
 \nonumber \\
&& + \bra f_1 |z\uparrow\ket ^{*}\bra f_2 
|z\uparrow\ket ^{*}\bra f_1 |z\downarrow\ket \bra f_2 
|z\downarrow\ket )] + \xi t_{extra}
\eeqa
where $ t_{extra}$ is
\beqa
  && { 1 \over 2}(|\bra f_1 |z\uparrow\ket \bra f_2
 |{z\uparrow}\ket |^{2} 
  +|\bra f_1 |z\downarrow\ket \bra f_2 |z\downarrow\ket |^{2}  
\nonumber \\
  && -\bra f_1 |z\uparrow\ket \bra f_2 |z\uparrow\ket 
 \bra f_1| z\downarrow\ket ^{*}
  \bra f_2| z\downarrow\ket ^{*} 
  - \bra f_1 |z\uparrow\ket ^{*}\bra f_2 |z\uparrow\ket^{*}\bra f_1| 
  z\downarrow\ket \bra f_2| z\downarrow\ket  
\nonumber \\  
  && -|\bra f_1 |z\uparrow\ket \bra f_2 |z\downarrow\ket |^{2} 
  - \bra f_1 |z\downarrow\ket \bra f_2 |z\uparrow\ket |^{2}  
\nonumber \\
  && - \bra f_1 |z\uparrow\ket \bra f_2 |z\downarrow\ket  
  \bra f_1| z\downarrow\ket ^{*}
  \bra f_2|z\uparrow \ket ^{*}  
  - \bra f_1 |z\downarrow\ket \bra f_2 |z\uparrow\ket  
  \bra f_1| z\downarrow\ket ^{*}
  \bra f_2| z\uparrow\ket ^{*})
\eeqa
In the first two terms in  $ t_{extra}$,
the transformation to z quantization axis
induced a probability for finding two  electrons spin up of order of $\xi$.
In this case these two terms 
violates total angular momentum conservation. 
And therefore the decoherence parameter
 $\xi$ 
for two spin correlated electrons
has to be zero. Otherwise not only the foundations of quantum mechanics
such as equation (\ref{spinSL}), becomes not true but also
 uniformity of space
will be lost as soon as $\xi \ne 0$ takes place.  
The decoherence or separation of two correlated
spin  ${1 \over 2}$  particles in the form of equation (\ref {spin})
could not occur \cite{WCZ} without violating
 the well established laws of physics.

\section{Concluding remarks}
Our naive intuition cultivated from interference 
phenomena of photons and electrons 
tells us that the interference terms in probabilities  
are 
one of the typical quantum mechanical quantities and it 
is natural to think that 
 a break down of
 quantum mechanics might be accounted  
by changing the interference terms.
 However that is not so for the correlated two neutral kaon case:
Because 
even if the interference term is zero in a given bases, it could be large
 in another bases as seen in the example given 
in section 2.
In other words, even if the probability of a system is described only 
 by a sum
 of probabilities (no interference terms),  it dose not imply that
 the system is ``more classical'' or separable or disentangled.
Another example for which our cultivated intuition for quantum mechanics
 doesn't work
is exhibited\cite{Hank} in  neutrino oscillations: 
Our naive intuition predicts that
the smaller the mass difference(the gap of two energy levels)
, the larger the rate of transition
between them.  However it is the  other way around; we observe easily
(transition happens more often) for the larger mass difference
 as the probability
of finding ${\nu}_{l}$ types neutrino 
in a beam which is initially pure $\nu$ neutrino, 
as  function of time,
is given by
\beqa
P(\nu \rightarrow {\nu}_{l}) \sim sin^2({\Delta M L \over 4E})
\qquad (\nu \ne \nu_l)
\eeqa
where $\Delta M^{2}$ = $m^{2}_\nu$ - $m^{2}_{{\nu}_{l}}$
and E and L are the energy and  traveled distance of $\nu$.

We have shown that
the decoherence in correlated two neutral kaons in the form of
modified interference terms is limited ;  equation(\ref{decoh1})
 with a condition $|\zeta'| \le 10^{-2}$ 
 and  equation(\ref{phil})
 with a condition $|\zeta| \le 10^{-5}$ may be used 
to analyze experimental data involving measurement of
 two neutral kaons in  strangeness states,
$(K^{0}, {\bar K}^{0})$    and  in  mass eigenstates,
 $(K_{S}, K_{L})$ respectively.

\acknowledgments{
The author thanks M. Suzuki for 
comments, P. Eberhard for discussion,  and the
theory group for hospitality}

\end{document}